\documentclass[11pt, twoside, fleqn]{article}

\usepackage{a4wide}
\usepackage[fleqn]{amsmath}
\usepackage{amssymb, exscale}
\usepackage{epsfig}
\usepackage{titlesec}
\usepackage{cite}
\usepackage[marginal]{footmisc} 
\usepackage[overload]{textcase} 
\usepackage{subfigure}
\usepackage{color}

\usepackage{latexsym}
\usepackage{graphics}
\usepackage{pst-node}
\usepackage{epsf}
\usepackage{exscale}

\usepackage[singlelinecheck=false,margin=\parindent,labelfont={small,sc},textfont={small,sf}]{caption}

\allowdisplaybreaks[1]

\newfont{\bigrm}{cmr10 at 12pt}
\newfont{\Bigrm}{cmr12 at 16pt}
\newfont{\hermes}{cmtt10 at 10pt}

\makeatletter

\renewcommand\thesection{\oldstylenums{\arabic{section}}}
\renewcommand\thesubsection{\thesection.\oldstylenums{\arabic{subsection}}}
\renewcommand\thesubsubsection{\thesubsection.\oldstylenums{\arabic{subsubsection}}}

\@addtoreset{equation}{section}

\newcommand{\sectionname}{Section}
\newcommand{\marktotoc}[1]{\renewcommand{\sectionname}{#1}}

\newcommand\secformat[1]{\sc{\MakeTextUppercase{#1}}}
\newcommand\subsecformat[1]{\sc{\MakeTextLowercase{#1}}}
\newcommand\subsubsecformat[1]{{\it #1}}

\titleformat{\section}[block]{}{\thesection\quad}{1em}{\secformat}[]
\titleformat{\subsection}[block]{}{\thesubsection\quad}{1em}{\subsecformat}[]
\titleformat{\subsubsection}[block]{}{\thesubsubsection\quad}{1em}{\subsubsecformat}[]

\titlespacing*{\section}{0pt}{2\baselineskip}{2\baselineskip}
\titlespacing*{\subsection}{0pt}{\baselineskip}{\baselineskip}
\titlespacing*{\subsubsection}{0pt}{\baselineskip}{\baselineskip}

\makeatother

\newcommand{\relsp}[1]{\mathrel{\phantom{{#1}}}{}}
\newcommand{\binsp}[1]{\mathbin{\phantom{{{}#1{}}}}}

\newcommand{\diag}{\mathrm{diag}}

\newcommand{\SU}{\mathrm{SU}}

\newcommand{\Lr}[1]{L^{\mathrm{r}}_{#1}}

\newcommand{\kr}[1]{k^\mathrm{r}_{#1}}
\newcommand{\Kr}[1]{K^\mathrm{r}_{#1}}

\newcommand{\bra}{\langle}
\newcommand{\ket}{\rangle}
\renewcommand{\d}{\partial}
\newcommand{\dd}{\mathrm{d}}
\newcommand{\ord}[1]{O\!\left(#1\right)}
\newcommand{\comm}[2]{[#1,#2]}
\newcommand{\anticomm}[2]{\{#1,#2\}}

\newcommand{\ri}{\mathrm{i}}

\newcommand{\fs}{\; .}

\newcommand{\co}{\; ,}

\newcommand{\eom}{\sc eom\rm}
\newcommand{\lecs}{\sc lec\rm s}
\newcommand{\lec}{\sc lec\rm}
\newcommand{\chpt}{$\chi$\sc{pt}\rm}
\newcommand{\cpt}[1]{$\chi$\sc pt\rm${}^\gamma_{#1}$}

\newcommand{\qcd}{\sc qcd\rm}

\newcommand{\nlo}{\sc nlo\rm}

\newcommand{\Mk}{M_K}

\begin{document}

\thispagestyle{empty}

\begin{flushright}
IFIC/07--53
\vspace{\baselineskip}\par
\end{flushright}


\vspace*{0.8cm}
\begin{center}
{\Bigrm Electromagnetic low--energy constants in {\Large \chpt}}
\vskip 0.6cm
{\bigrm Christoph Haefeli$^1$, 
Mikhail A.~Ivanov$^2$, and
Martin Schmid$^3$ }
\vskip 3ex
{\bigrm\it 
$^1$Departament de F\'{\i}sica Te\`orica,
IFIC, Universitat de Val\`encia -- CSIC,}
\\
{\bigrm\it Apt. Correus 22085, E--46071 Val\`encia, Spain}
\\[2ex]
{\bigrm\it $^2$Laboratory of Theoretical Physics,
Joint Institute for Nuclear Research,}
\\
{\bigrm\it 141980 Dubna (Moscow region), Russia}
\\[2ex]
{\bigrm\it $^3$Institute for Theoretical Physics, University of Bern,}
\\
{\bigrm\it Sidlerstr. 5, 3012 Bern,  Switzerland } 
\end{center}

\vskip 0.8cm
\hrule
\vskip 2.0ex
{\small
\setlength{\baselineskip}{14pt}
\noindent
{\bf Abstract}
\vskip 1.0ex\noindent
We investigate three--flavour chiral perturbation theory including virtual
photons in a limit where the strange quark mass is much larger than the
external momenta and the up and down quark masses, and where the external
fields are those of two--flavour chiral perturbation theory. In particular we
work out the strange quark mass dependence of the electromagnetic two--flavour 
low--energy constants $\,C$ and $k_i$. We expect that these relations will be
useful for a more precise determination of the electromagnetic low--energy
constants. 
}
\vskip 2.0ex
\hrule
\vspace*{1cm}


\noindent
\sc{pacs}\rm: 11.30.Rd{}, {12.39.Fe}{},  {13.40.Dk}{},  {13.40.Ks}{}

\setlength{\baselineskip}{17pt}

\newpage

\section{Introduction}

\noindent
Chiral perturbation theory (\chpt) \cite{Weinberg,GL:Ann,GL:NPB} is the
effective theory of \qcd{} at low energies. It relies on an effective
Lagrangian whose coupling constants are the chiral low--energy constants
(\lecs{}). They are independent of the light quark masses and encode the
influence of the heavy degrees of freedom that are not contained in the
Lagrangian explicitly. For many phenomenological applications the predictivity
of \chpt{} depends on realistic estimates of these \lecs{}. An up--to--date
account of our knowledge about the \lecs{} can be found in the recent
conference reports of Ecker \cite{Ecker:lecs} and Bijnens \cite{Bijnens:lecs}.

In this article we are concerned with the electromagnetic \lecs{} of
\chpt{} in the (natural parity) meson sector including virtual photons. 
In the following we
abbreviate the effective theory with three flavours with
\cpt{3}~\cite{Urech,Neufeld}, and
accordingly  for two flavours with
\cpt{2}~\cite{Knecht.Urech,Steininger}. Recently, quite some progress has 
been achieved estimating the \nlo{} electromagnetic \lecs{} in \cpt{3}
\cite{Anant.Mouss,Baur.Urech,Bijnens.Prades,MoussallamGauge,K_i}, while
little is known about the pertinent \lecs{} in \cpt{2}, to the
best of our knowledge.  
In such a situation the following strategy may be pursued
\cite{Gasser:atom,Sazdjian-I}:
if one limits the external momenta to values small compared to the kaon and
eta mass and treats $m_u,m_d$ as small in comparison to $m_s$
\begin{equation}
  |p^2| \ll M_K^2 \co \qquad m_u,m_d \ll m_s \co
\end{equation}
the degrees of freedom of the kaons and the eta freeze. In this region one may work
out relations among the \lecs{} in \cpt{2} and \cpt{3}, which allow one to
estimate the electromagnetic \lecs{} in \cpt{2} through the knowledge of the
ones in \cpt{3}. 
The purpose of the present article is to
systematically provide \emph{all} relations between the
$\ord{e^2,e^2 p^2,e^4}$ 
\lecs{} in \cpt{2} and \cpt{3} at one--loop order. The calculation is performed along
the lines outlined in \cite{Nyffeler,Schmid:PhD,matchingI}. Briefly, the
method consists in a non--trivial matching between the three--flavour versus
the two--flavour generating functional of the effective theory.
\footnote{After we had performed these matching relations, we were informed
  by Marc Knecht that the relations at $\ord{e^2}$ and $\ord{e^2p^2}$ had also
  been derived by Nehme \cite{NehmePhD}. We agree with the
  relations given there.} 

We briefly comment on related work in the literature. Analogous relations
between the two--flavour and the three--flavour \lecs{} in the strong sector
have been provided by Gasser and
Leutwyler in \cite{GL:NPB}. Recently, we have worked out the same relations to
the next higher order (at two--loops) in the perturbative expansion
\cite{matchingI}, see also \cite{Mouss:Sigma,Schweizer:B} for earlier
contributions of such relations at two--loops. Finally, analogous work was
performed at one--loop accuracy in the baryonic sector in \cite{meissner_frink}.

The remainder of the article is organised as follows. After setting the
notation in Sec.~\ref{sec:ChPT_photons}, we give some details on the
derivation of the matching relations in Sec.~\ref{sec:method}. 
Section~\ref{sec:numerics}  contains a numerical analysis of
the matching relations, leading to estimates of the electromagnetic \nlo{}
\lecs{} in \cpt{2}.


\section{ Including virtual photons in  \chpt{}}
\label{sec:ChPT_photons}

A general procedure to construct the effective theory with photons in the
mesonic sector for three light flavours (\cpt{3}) has been proposed by Urech
\cite{Urech}. The two--flavour effective theory \cpt{2} may
be constructed along the same lines \cite{Knecht.Urech,Steininger}. We set our
notation, following the nomenclature of the \lecs{}
introduced by Urech \cite{Urech} for \cpt{3}, and by Knecht and Urech
\cite{Knecht.Urech} for \cpt{2}.\\ 

\noindent
The basic building block of the chiral Lagrangian is the Goldstone matrix
field  $u(\phi)$ which transforms under a chiral rotation $g=(g_L,g_R)\in
\SU(n)\times\SU(n)$ as 
\[u(\phi)\stackrel{g}{\longrightarrow} u(\phi')=g_Ru(\phi)h(g,\phi)^{-1} =
h(g,\phi)u(\phi)g_L^{-1}\co\]
where $h$ is called the compensator field. The mesonic Lagrangian then consists
of operators $X$ that either transform as
\begin{equation}
\label{eq:chirtrans}
X\stackrel{g}{\longrightarrow} h(g,\phi)Xh(g,\phi)^{-1}\co
\end{equation}
or remain invariant under chiral transformations. As a result, (products of)
traces of products of chiral operators $X$ are chiral invariant. 
The elementary building blocks of the effective Lagrangian that have the
transformation property of Eq.~(\ref{eq:chirtrans}) and furthermore contain 
the external vector $v_\mu$, axial $a_\mu$ (both traceless), scalar $s$, and
pseudoscalar $p$ sources are given by
\begin{equation}
\label{eq:buildblockI}
\begin{split}
  u_\mu &= \ri\left[u^\dag (\d_\mu-\ri r_\mu)u-u (\d_\mu-\ri
    l_\mu)u^\dag\right]
\co\\
  \chi_\pm &= u^\dag\chi u^\dag\pm u\chi^\dag u\co
\end{split}
\end{equation}
where 
\begin{equation}
  \begin{split}
  r_\mu &= v_\mu+a_\mu+Q_RA_\mu\co\qquad l_\mu = v_\mu-a_\mu+Q_LA_\mu\co\\
  \chi &= 2B_0(s+ip)\co
  \end{split}
\end{equation}
with $A_\mu$ the photon field and $Q_{L,R}$ spurion sources with the
transformation properties
\begin{align*}
  Q_R\stackrel{g}{\longrightarrow} g_R\,Q_R\,g_R^\dag\co\qquad
  Q_L\stackrel{g}{\longrightarrow} g_L\,Q_L\,g_L^\dag\fs 
\end{align*}
They are also contained in the building blocks 
\begin{align}
  q_R = u^\dag Q_R u\co\qquad q_L=u Q_L u^\dag\co
\end{align}
which transform according to Eq.~(\ref{eq:chirtrans}).
Below, we will consider 
constant sources $Q_R = Q_L = Q$ only, 
and of phenomenological interest are the cases with two $(Q=Q_2)$, as well as
three light flavours $(Q=Q_3)$,
\begin{equation}
\label{eq:charge}
 Q_2=\tfrac{e}{3}\diag(2,-1)\co\qquad Q_3=\tfrac{e}{3}\diag(2,-1,-1)\fs
\end{equation}
Note that $Q_3$ is traceless, while $Q_2$ is not. For three flavours the
leading order Lagrangian reads in Euclidean space--time, 
\begin{equation}
  \begin{split}
    \mathcal{L}_2^{(3)} &= \frac{F_0^2}{4}\bra u\cdot u-\chi_+\ket-C_0\bra q_Lq_R\ket
    +\tfrac{1}{4}F_{\mu\nu}F_{\mu\nu}+\tfrac{1}{2}(\d_\mu A_\mu)^2
    \co
\end{split}
\end{equation}
where the superscript ${(3)}$ labels the number of flavours. Further, 
$u\cdot u \equiv u_\mu u_\mu$, $F_{\mu\nu} = \d_\mu A_\nu - \d_\nu A_\mu$
is the fieldstrength of the photon, and the gauge fixing term is put in the
Feynman gauge, as is customary in \cpt{2,3}. 
The symbol $\bra\cdot\ket$ denotes the trace of the flavour matrix enclosed.  
For two flavours the leading order Lagrangian amounts to have the same form
alike for three flavours, with the difference of restricting the $u$ fields to
elements of $\SU(2)$, and similar for the sources. Furthermore, the \lecs{}
$F_0$, $B_0$, and $C_0$ are to be replaced with $F$, $B$, and $C$,
respectively. To distinguish two-- from  three--flavour fields,
we decorate the former ones with a superscript $^\pi$. In summary, 
\begin{equation}
  \begin{split}
    \label{eq:lagr2}
    \mathcal{L}_2^{(2)} &= \frac{F^2}{4}\bra u^\pi\cdot u^\pi-\chi_+^\pi\ket
    -C \bra q_L^\pi q_R^\pi \ket
    +\tfrac{1}{4}F^\pi_{\mu\nu}F^\pi_{\mu\nu}
    +\tfrac{1}{2}(\d_\mu A^\pi_\mu)^2
    \fs
\end{split}
\end{equation}
%
%
\begin{table}[!ht]
\begin{center}
\def\arraystretch{1.5}
\begin{tabular}{rll}
\hline\hline
$j$ & $\binsp{-}x_j$ \hspace*{3cm} & $\binsp{-}X_j$ \\
\hline\\[-3ex]
 1  
& 
{} $ -\tfrac{1}{4}\bra u^\pi\cdot u^\pi\ket^2 $
&
{} $ -\bra u\cdot u\ket^2 $
\\
2
&
{} $ -\tfrac{1}{4}\bra u^\pi_\mu u^\pi_\nu\ket^2 $
&
{} $ -\bra u_\mu u_\nu\ket^2 $
\\
3
&
{} $ -\tfrac{1}{16}\bra\chi^\pi_+\ket^2 $
&
{} $ -\bra(u\cdot u)^2\ket $
\\
4
&
$ \binsp{-} \tfrac{i}{4}\bra u^\pi_\mu\chi^\pi_{-\mu}\ket $
&
$ \binsp{-} \bra u\cdot u\ket\bra\chi_+\ket $
\\
5
&
$ \binsp{-} \tfrac{1}{2}\bra {f^\pi_-}^2\ket $
&
$ \binsp{-} \bra u\cdot u\,\chi_+\ket $
\\
6
&
{} $ -\tfrac{i}{4}\bra f^\pi_{+\mu\nu}[\,u^\pi_\mu\,,\,u^\pi_\nu\,]\ket $
&
{} $ -\bra\chi_+\ket^2 $
\\
7
&
$ \binsp{-} \tfrac{1}{16}\bra\chi^\pi_-\ket^2 $
&
{} $ -\bra\chi_-\ket^2 $
\\
8
&
{} $ -\tfrac{1}{8}(\det\chi^\pi_++\det\chi^\pi_-) $
&
{} $ -\tfrac{1}{2}\bra\chi_+^2+\chi_-^2\ket $
\\
9
&
$ \binsp{-} \bra {\tilde{f}_+^{\pi\,2}}+{f^\pi_-}^2\ket $
&
$ \binsp{-} \tfrac{i}{2}\bra f_{+\mu\nu}[\,u_\mu\,,\,u_\nu\,]\ket $
\\
10
&
{} $ -\tfrac{1}{16}\bra{\chi^\pi_+}^2-{\chi^\pi_-}^2\ket $
&
{} $ -\tfrac{1}{4}\bra f_+^2-f_-^2\ket $
\\
11
&
{} $ -\tfrac{1}{4}\bra f_{+\mu\nu}^\pi\ket^2 $
&
{} $ -\tfrac{1}{2}\bra f_+^2+f_-^2\ket $
\\
12
&
&
{} $ -\tfrac{1}{4}\bra\chi_+^2-\chi_-^2\ket $
\\[1ex]
\hline\hline\\[-1ex]
\end{tabular}
\caption{
\setlength{\baselineskip}{14pt}
Basis of the strong operators at order $p^4$ 
in Euclidean metric for two $(x_j)$ and for three flavours $(X_j)$.}
\label{tab:x_j} 
\end{center}
\end{table}
For the \nlo{} Lagrangian $\mathcal{L}_4$ we need the following additional
building blocks:
\begin{equation}
\label{eq:buildblockII}
  \begin{split}
    f_{\pm\mu\nu}&=ul_{\mu\nu}u^\dag\pm u^\dag r_{\mu\nu}u\co\\
    \chi_{\pm\mu}&=\nabla_\mu\chi_\pm-\tfrac{\ri}{2}\anticomm{\chi_\mp}{u_\mu}
\co\\
    q_{R\mu} & = \nabla_\mu q_R - \tfrac{\ri}{2}\comm{u_\mu}{q_R}\co\\
    q_{L\mu} & = \nabla_\mu q_L + \tfrac{\ri}{2}\comm{u_\mu}{q_L}\co
  \end{split}
\end{equation}
where we have introduced the field strengths
\begin{align*}
    y_{\mu\nu}&=\d_\mu y_\nu-\d_\nu y_\mu-\ri\comm{y_\mu}{y_\nu}
\co\qquad
y\,\in\, \{r,l\}
\co
\end{align*}
and the covariant derivative $\nabla_\mu$ in terms of the chiral connection
$\Gamma_\mu$,
\begin{align*}
  \nabla_\mu X&=\d_\mu X+[\Gamma_\mu,X]\co\\ 
  \Gamma_\mu&=\tfrac{1}{2}\left[u^\dag(\d_\mu-\ri r_\mu)u
  +u(\d_\mu-\ri l_\mu)u^\dag\right]\fs
\end{align*}
It is worth noting that $f_{+\mu\nu}$ for two--flavours is not traceless, since
the charge matrix $Q_2$ is not. To be in line with the basis operators
introduced by Gasser and Leutwyler \cite{GL:Ann}, it is convenient to
introduce in addition the traceless operator $\tilde{f}_{+\mu\nu}^\pi$,
\begin{equation}
\tilde{f}_{+\mu\nu}^\pi=f_{+\mu\nu}^\pi-\tfrac{1}{2}\bra f_{+\mu\nu}^\pi\ket
\fs
\end{equation}
The \nlo{} Lagrangians $\mathcal{L}_4$ then read
\begin{align}
  \mathcal{L}_4^{(2)} &= \sum_{j=1}^{11}l_jx_j+F^2\sum_{j=1}^{11}k_j w_j
+F^4\sum_{j=12}^{14}k_j w_j\co\\
  \mathcal{L}_4^{(3)} &= \sum_{j=1}^{12}L_jX_j+F_0^2\sum_{j=1}^{14}K_j W_j
+F_0^4\sum_{j=15}^{17}K_j W_j 
\fs
\end{align}
In the following, we call the
operators $x_j,X_j$ and the \lecs{} $l_j,L_j$ \emph{strong operators} and
\emph{strong \lecs{}}, as these operators do not vanish when switching off the
electromagnetic coupling constant, with the exception of
$x_{11}$. Accordingly, we denote the operators/\lecs{}
$w_j,W_j/k_j,K_j$ \emph{electromagnetic operators/\lecs{}}. The strong
operators $x_j,X_j$ have been introduced by Gasser and Leutwyler
\cite{GL:Ann,GL:NPB} and for convenience, we reproduce them here in
Tab.~\ref{tab:x_j}. The coefficients  $h_i$ and $H_i$ of the contact terms --
introduced by Gasser and Leutwyler in \cite{GL:Ann,GL:NPB} -- are related to
our \lecs{} as
\begin{equation}
  \label{eq:contact}
  \begin{split}
    h_1-h_3 &= l_8\co&\qquad h_2 &= l_9\co\qquad
    h_1+h_3=l_{10}\co\\ 
    H_1&=L_{11}\co&\qquad H_2&=L_{12}\fs
  \end{split}
\end{equation}

%
\begin{table}[!h]
\begin{center}
\def\arraystretch{1.5}
\begin{tabular}{rll}
\hline\hline
$j$ & $\binsp{-}w_j$ \hspace*{5cm} & $\binsp{-}W_j$ \\
\hline\\[-3ex]
 1  &  $ \binsp{-}\tfrac{1}{2}\bra u^\pi\!\cdot u^\pi\ket\bra
 q^\pi_R{}^2+q^\pi_L{}^2\ket $  &
       $ \binsp{-}\tfrac{1}{2}\bra u\!\cdot u\ket\bra q_R^2+q_L^2\ket $ \\

 2  &  $ \binsp{-}\bra u^\pi\!\cdot u^\pi\ket\bra q^\pi_R q^\pi_L\ket $  &
       $ \binsp{-}\bra u\!\cdot u\ket\bra q_R q_L\ket $ \\

 3  &  ${}-\bra u^\pi_\mu q^\pi_R\ket^2-\bra u^\pi_\mu q^\pi_L\ket^2 $ &
       ${}-\bra u_\mu q_R\ket^2-\bra u_\mu q_L\ket^2 $ \\

 4  &  $ \binsp{-}\bra u^\pi_\mu q^\pi_R\ket\bra u^\pi_\mu q^\pi_L\ket $ &
       $ \binsp{-}\bra u_\mu q_R\ket\bra u_\mu q_L\ket $ \\

 5  &  $ {}-\tfrac{1}{2}\bra\chi^\pi_+\ket\bra q^\pi_R{}^2+q^\pi_L{}^2\ket $ &
       $ \binsp{-}\bra u\!\cdot u (q_R^2+q_L^2)\ket $ \\  

 6  &  ${}-\bra\chi^\pi_+\ket\bra q^\pi_R q^\pi_L\ket$ &
       $ \binsp{-}\bra u\!\cdot u\anticomm{q_R}{q_L}\ket $  \\

 7  &  ${}-\tfrac{1}{2}\bra\chi^\pi_+(q^\pi_R+q^\pi_L)\ket\bra
 q^\pi_R+q^\pi_L\ket $ &  
       ${}-\tfrac{1}{2}\bra\chi_+\ket\bra q_R^2+q_L^2\ket $ \\ 

 8  &  ${}-\bra\chi^\pi_-\comm{q^\pi_R}{q^\pi_L}\ket $ &
       ${}-\bra\chi_+\ket\bra q_R q_L\ket $ \\

 9  &  $ \binsp{-}\ri\bra
 u^\pi_\mu(\comm{q^\pi_{R\mu}}{q^\pi_R}-\comm{q^\pi_{L\mu}}{q^\pi_L})\ket $ &  
       ${}-\bra\chi_+(q_R^2+q_L^2)\ket $ \\

 10 &  $ \binsp{-}\bra q^\pi_{R\mu}q^\pi_{L\mu}\ket $ &
       ${}-\bra\chi_+\anticomm{q_R}{q_L}\ket  $ \\

 11 &  $ \binsp{-}\bra q^\pi_R\!\cdot q^\pi_R+q^\pi_L\!\cdot q^\pi_L\ket $ &
       ${}-\bra\chi_-\comm{q_R}{q_L}\ket $ \\ 

 12 &  ${}-\tfrac{1}{4}\bra q^\pi_R{}^2+q^\pi_L{}^2\ket^2 $ &
       $ \binsp{-}\ri\bra u_\mu(\comm{q_{R\mu}}{q_R}-\comm{q_{L\mu}}{q_L})\ket
       $ \\ 

 13 &  ${}-\tfrac{1}{2}\bra q^\pi_R q^\pi_L\ket\bra
 q^\pi_R{}^2+q^\pi_L{}^2\ket $ &  
       $ \binsp{-}\bra q_{R\mu}q_{L\mu}\ket $ \\

 14 &  ${}-\bra q^\pi_R q^\pi_L\ket^2 $ & 
       $ \binsp{-}\bra q_R\!\cdot q_R+q_L\!\cdot q_L\ket $ \\

 15 & & $ {}-\bra q_R q_L\ket^2 $ \\

 16 & & $ {}-\tfrac{1}{2}\bra q_R q_L\ket\bra q_R^2+q_L^2\ket $ \\

 17 & & $ {}-\tfrac{1}{4}\bra q_R^2+q_L^2\ket^2 $\\[1ex]
\hline\hline\\[-1ex]
\end{tabular}
\caption{
\setlength{\baselineskip}{14pt}
Basis of the electromagnetic operators at order $e^2p^2$ and $e^4$ 
in Euclidean metric for two $(w_j)$ and for three flavours $(W_j)$.
The abbreviation $q_A\!\cdot q_A = q_{A\mu}q_{A\mu}$ for
$A\in\{R,L\}$ is adopted.}
\label{tab:w_j} 
\end{center}
\end{table}
\noindent
We are using the set of the \nlo{} electromagnetic operators from Knecht and
Urech \cite{Knecht.Urech} for two, and from Urech 
\cite{Urech} for three flavours, cf. Tab.~\ref{tab:w_j}. 
We do not consider the odd--intrinsic parity
sector which accounts for the axial anomaly, see
e.g.~\cite{WZW,Kaiser,MoussAnomaly}. 
For the counting we rely on the standard \cpt{} assignment 
with $e^2 \sim{} \ord{p^2}$ and we make use of the
convention to write $\ord{e^4,e^2p^2,p^4}$ as $\ord{p^4}$, and similarly for
the Landau symbol at order $p^6$. 
\clearpage


\section{Integrating out the strange quark}
\label{sec:method}

This section is devoted to give some details on the derivation of the main
results presented below in Eq.~(\ref{eq:results}). We will follow the steps
outlined in \cite{Nyffeler,Schmid:PhD}.


\subsection{Generating functional}
\label{sec:genfunc}

We start by considering the generating functional $Z$ of \cpt{3}
\cite{GL:NPB}, 
\begin{align}
  \mathrm{e}^{- Z[v,a,s,p,Q_{L,R}]} &= 
  \mathcal{N} \int[\dd u][\dd A_\mu]\, \mathrm{e}^{ - \int \dd^d x\,
    \mathcal{L}_{\mathrm{eff}}^{(3)} }
  \co\\[1ex]
  \mathcal{L}_{\mathrm{eff}}^{(3)} &= \mathcal{L}_2^{(3)} + \mathcal{L}_4^{(3)} + \ldots
  \fs
\end{align}
It may be evaluated in a low--energy
expansion in the number of loops,
\begin{equation}
  Z = Z_0 + Z_1 + \ldots
  \co
\end{equation}
where $Z_0$ $(Z_1)$ collects the tree--level (one--loop)
contributions.  They are given by
\begin{align}
  \label{eq:z0}
  Z_0 &= \bar{S}_2 \co
  \\
  \label{eq:z1}
  Z_1 &= \bar{S}_4 + \tfrac{1}{2}\ln\frac{\det D}{\det D^0}
  \co
\end{align}
where $\bar{S}_n$ denotes the classical action
\begin{equation}
\label{eq:barSn}
  \bar{S}_n = \int \dd^d x\, 
  \mathcal{L}_n^{(3)}(u^\text{cl},A^\text{cl},v,a,s,p,Q_{L,R})
  \co
\end{equation}
the Goldstone Boson fields $u^\text{cl}$ and the photon field $A^\text{cl}_\mu$ being evaluated
at the solution of the classical equation of motion (\eom{}), 
\begin{equation}
  \label{eq:EOM}
  \begin{split}
    \nabla_\mu
    u^\text{cl}_\mu+\tfrac{\ri}{2}\tilde{\chi}^\text{cl}_-+2\ri\tfrac{C_0}{F_0^2}\comm{q_R^\text{cl}}{q_L^\text{cl}}&=0\co\\
    \Delta A^\text{cl}_\mu-\tfrac{F_0^2}{2}\bra u^\text{cl}_\mu(q_R^\text{cl}-q_L^\text{cl})\ket&=0\co
  \end{split}
\end{equation}
where $\tilde{\chi}_-^\text{cl}$ denotes the traceless part of $\chi_-^\text{cl}\fs$
The Green's function of the differential operator $D$ [$D^0$] of
Eq.~(\ref{eq:z1}) is the [free] propagator $G(x,y)$\,$[G_0(x,y)]$,
\begin{equation}
 D_{AC}(x) G_{CB}(x,y) = \delta_{AB}\, \delta^{(d)}(x-y)
\fs
\end{equation}
The explicit form of $D$ was first given by Urech\footnote{Capital flavour indices
  $A,B,C,\ldots$ run from $1,\ldots,12$, lower case flavour indices from
  $1,\ldots,8$, they span the meson flavour space, and greek indices
  $\rho,\sigma,\ldots$ from $1,\ldots,4$ for the photon field components. The
  symbols $\lambda^a$ stand for the Gell--Mann matrices.} \cite{Urech},  
\begin{equation}
  \label{eq:diffop}
  D(x)=-\Sigma^2(x)+\Lambda(x)\co\qquad \Sigma_\mu(x)=\d_\mu^x+Y_\mu(x)\co
\end{equation}
with\footnote{To ease notation, we will drop
  from now on the label $_\text{cl}$ for the fields that satisfy the \eom{}.}
\begin{equation}
\begin{split}
\label{eq:loopvert}
  Y_\mu &=
  \begin{pmatrix}
    \hat\Gamma_\mu^{ab} & X_\mu^{a\rho}\\
    X_\mu^{\sigma b} & 0
  \end{pmatrix},\qquad\Lambda=
  \begin{pmatrix}
    \sigma^{ab} & \tfrac{1}{2}\gamma^{a\rho}\\
    \tfrac{1}{2}\gamma^{\sigma b} & \rho\delta^{\sigma\rho}
  \end{pmatrix}\co\\[1ex]
  \hat{\Gamma}_\mu^{ab}&=-\tfrac{1}{2}\bra\comm{\lambda^a}{\lambda^b}\Gamma_\mu\ket\co\\
  X_\mu^{a\rho} &= -X^{\rho a}_\mu =
  -\tfrac{F_0}{4}\delta_\mu^\rho\bra\lambda^a(q_R-q_L)\ket\co\\[1ex]
    \sigma^{ab}&=\binsp{+}\tfrac{1}{8}\bra\comm{\lambda^a}{u_\mu}\comm{\lambda^b}{u_\mu}\ket+\tfrac{1}{8}\bra\anticomm{\lambda^a}{\lambda^b}\chi_+\ket\\
    &\relsp{=}-\tfrac{C_0}{4F_0^2}\bra\big(\comm{\lambda^a}{q_R+q_L}\comm{\lambda^b}{q_R+q_L}
 -\comm{\lambda^a}{q_R-q_L}\comm{\lambda^b}{q_R-q_L}\big)\ket\\
    &\relsp{=}-\tfrac{F_0^2}{4}\bra\lambda^a(q_R-q_L)\ket\bra\lambda^b(q_R-q_L)\ket\co\\[1ex]
    \gamma^a_\mu &=
    \tfrac{F_0}{2}\bra\lambda^a\left\{\nabla_\mu(q_R-q_L)+i\comm{u_\mu}{q_R+q_L}\right\}\ket\co\\
    \rho &= \tfrac{3}{8}F_0^2\bra(q_R-q_L)^2\ket\fs
\end{split}
\end{equation}
\noindent
The generating functional for two flavours $z$ is defined analogously
to the one with three flavours. For later purposes we explicitly introduce its
low--energy expansion up to one--loop,
\begin{equation}
  z = \bar{s}_2 + \bar{s}_4
   + \tfrac{1}{2}\ln\frac{\det d}{\det d^{\,0}}
   + \ldots
\co
\end{equation}
where $\bar{s}_{2,4}$ and the operator $d$ are the two--flavour equivalent of
Eq.~(\ref{eq:barSn}) and Eq.~(\ref{eq:diffop}), and the ellipsis stand for
two--loop corrections and higher. 
For a state-of-the-art evaluation of the two--flavour functional and more
details we refer the reader to Schweizer \cite{Schweizer}.


\subsection{matching}

We impose now the following
constraints on the three--flavour functional:
\begin{itemize}
\item[i)]  the external sources of \cpt{3} are restricted to the two--flavour
  subspace. The generating functionals shall depend on the \emph{same}
  external sources;
\item[ii)] $m_{u,d} \ll m_s$,\\
since the \lecs{} of \cpt{2,3} are independent of $m_{u,d}$, 
we will work in the chiral limit for the up and down quark
masses, i.e.~$m_{u,d}=0$, for simplicity;
\item[iii)] external momenta are restricted to values below the threshold of
  the massive fields, $|p^2| \ll M_K^2$.
\end{itemize}
We will refer to the limit that satisfies i), ii) and iii) as \emph{the
two--flavour limit} of the three--flavour theory. In this limit the
three--flavour functional reduces to the two--flavour functional, i.e. both
theories yield the same Green's functions in the low--energy region, 
\begin{equation}
\label{eq:match_genF}
  Z = z
  \co
\end{equation}
provided the \lecs{} of both theories are accordingly matched. 
In the following, we will solve this equation for the \lecs{}. Both sides
receive non--local contributions that are associated to the propagation of
massless pions and photons. Once the matching is fully worked out, these contributions
cancel against each other. At order $p^4$, this has  
been discussed in detail by Nyffeler and Schenk
\cite{Nyffeler} and further details shall also be given elsewhere 
\cite{matchingII}. To find the relations among the \lecs{} it suffices 
therefore to work out the \emph{local} parts of the generating
functionals. At order $p^4$ we have in the two--flavour limit, 
\begin{equation}
  \label{eq:match_genF2}
  \bar{S}_2 + \bar{S}_4
  + \tfrac{1}{2}\ln\frac{\det D}{\det D^0}\Big|_\text{local} = 
  \bar{s}_2 + \bar{s}_4
\fs
\end{equation}
The lhs. of Eq.~(\ref{eq:match_genF2}) is now being worked out. 
We start with the tree--level
contributions, and proceed with the one--loop corrections.


\subsubsection{Tree level: solution of the {\footnotesize EOM} in the two--flavour limit}

In view of the Eqs.~(\ref{eq:z0}\,,\,\ref{eq:z1}) we need to solve the
\eom{}
in the two--flavour 
limit. Due to the absence of strangeness containing external sources
[restriction i)] as well as
strangeness conservation, the following ansatz 
for the Goldstone Boson fields will turn out to be fruitful, 
\begin{equation}
  \label{eq:solfields}
  {u}={u}^\pi \mathrm{e}^{\tfrac{\ri}{2F_0}{\eta}\lambda_8}
  \co
\end{equation}
where in $u^\pi$ only the pions contribute non--trivially. 
Below, we will frequently identify without further notice
$3\times 3$ matrices that have only non--vanishing elements in their upper
left $2\times 2$ block with the $2\times 2$ matrices from the two--flavour
theory. Inserting the ansatz Eq.~(\ref{eq:solfields}) into the building blocks
Eqs.~(\ref{eq:buildblockI}) yields
\begin{equation}
\label{eq:transl1}
\begin{split}
      {u}_\mu&={u}^\pi_\mu-\tfrac{1}{F_0}\lambda^8\d_\mu{\eta}
\co\\[1ex]
    q_{L,R} &= q_{L,R}^\pi-e_{33}\bra q_{L,R}^\pi\ket
\co\\
    { \chi}_\pm&= \tfrac{B_0}{B}\chi_\pm^\pi\cos\alpha
     - \ri\tfrac{B_0}{B}\chi_\mp^\pi\sin\alpha+4B_0m_se_{33}
    \begin{cases}
       \;\cos 2\alpha & \!\! (+) \\
      \ri\sin 2\alpha & \! \! (-)
    \end{cases}
\co\\
\alpha &= \eta/(\sqrt{3}F_0) 
\co\qquad
e_{33} = \text{diag}(0,0,1)
\fs
\end{split}
\end{equation}
And similar for the building blocks of
Eqs.~(\ref{eq:buildblockII}),
\begin{equation}
\label{eq:transl2}
\begin{split}
    {f}_{+\mu\nu}&= f^\pi_{+\mu\nu}-e_{33}\bra f^\pi_{+\mu\nu}\ket
\co\qquad
    &{f}_{-\mu\nu}= {f}^\pi_{-\mu\nu}
\co\\
    q_{A\mu} &= q_{A\mu}^\pi
\co\qquad 
&A\in\{R,L\}
\fs
\end{split}
\end{equation}
Next, we write down the \eom{} of the $\eta$ field, 
\begin{align}
\label{eq:eometa}
\left(\Delta-{M}_{\eta}^2\right){\eta}
&=
\frac{F_0}{4\sqrt{3}}\left[\tfrac{B_0}{B}\bra\chi_+^\pi\ket\sin\alpha
+\ri\tfrac{B_0}{B}\bra\chi_-^{\pi}\ket\cos\alpha
+8B_0m_s\sin 2\alpha\right]-{M}_\eta^2{\eta}
\co
\end{align}
where $M_\eta^2 = \tfrac{4}{3} B_0m_s$ is the eta mass squared at tree level
at $m_{u,d}=0$ [similarly we will use $M_K^2 = B_0m_s$ below]. The \eom{} 
may be solved recursively for small $\alpha$. Note that the sum of the last
two terms in Eq.~(\ref{eq:eometa}) is of order $\alpha^3$. The differential
equation suggests a counting in which every occurrence of an $\eta$ particle
counts as order $p^2$ in the two--flavour limit,
\begin{align}
  \label{eq:eta}
  \eta &= -\frac{\ri\sqrt{3}F_0}{16B m_s}\bra{\chi}^\pi_-\ket + \ord{p^4}
  \fs
\end{align}
As a result, we obtain a systematic low--energy expansion of the $\SU(3)$
building blocks. 
\begin{figure}[!b]
  \centering
  \epsfig{file=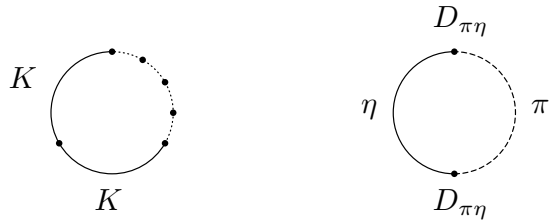,height=7\baselineskip,bbllx=129,bblly=634,bburx=300,bbury=716}
  \caption{
    \setlength{\baselineskip}{14pt}
    (Left:) 
    Diagrammatic illustration for a specific contribution to
    $\ln\det D_K$ in Eq.~(\ref{eq:det}). The vertices denote insertions from
    external fields, related to the operator $D$. Only kaons flow through the
    loop. The whole determinant
    consists of a sum of such diagrams, ordered by an increasing number of
    insertions, being equivalent to the low--energy expansion introduced in
    the text.
    (Right:)
    Pion--eta mixing at order $p^4$. In the two--flavour limit this diagram
    yields to local terms only.
  }
  \label{fig:det.mix}
\end{figure}

Before proceeding, we add a remark: to be precise, the pions of
\cpt{3} differ from their two--flavour equivalent, since
they satisfy different \eom{}s. Indeed, in the two--flavour limit, we find
\begin{equation}
\label{eq:su3eom}
  \nabla_\mu
u^\pi_\mu+\tfrac{\ri}{2}\tfrac{B_0}{B}\tilde{\chi}^\pi_-\cos\alpha+\tfrac{1}{2}\tfrac{B_0}{B}\tilde{\chi}^\pi_+\sin\alpha+2\ri\tfrac{C_0}{F_0^2}\comm{q_R^\pi}{q_L^\pi}=0\co
\end{equation}
to be compared with the \eom{} of \cpt{2}, 
\begin{equation}
  \nabla_\mu
u^\pi_\mu+\tfrac{\ri}{2}\tilde{\chi}^\pi_-+2\ri\tfrac{C}{F^2}\comm{q_R^\pi}{q_L^\pi}=0\fs
\end{equation}
However, expanding the trigonometric functions in Eq.~(\ref{eq:su3eom}), one
observes that the difference is of order $p^4$; hence, it affects the matching
relations only beyond the accuracy we are working. Also at the level of
$\bar{S}_2$ -- which is the only one to matter for the \eom{} -- the \lecs{} of
both theories coincide due to the matching condition (\ref{eq:match_genF}). These are the reasons why we
do not distinguish between the pions of both theories in this article. However, if one wishes to
carry out the matching beyond one--loop order, this issue requires a
considerable deeper examination, see also  \cite{matchingII}. Similar remarks
apply for the photon.  

By now, it is straight forward to evaluate the tree level
diagrams of 
$\bar{S}_{2,4}$ in the two--flavour limit using
Eqs.~(\ref{eq:transl1}\,,\,\ref{eq:transl2}\,,\,\ref{eq:eta}). To avoid
overflowing formulae, for $\bar{S}_4$
we only show the reduction of the electromagnetic
operators  explicitly. We find,
\begin{equation}
\label{eq:treeResults}
\begin{split}
\bar{S}_2 &= 
\int\dd^d x\,\Big[
\tfrac{F_0^2}{4} \bra u^\pi\cdot u^\pi \ket
-\tfrac{F_0^2}{4} (B_0/B) \bra \chi_+^\pi \ket
-C_0 \bra q_L^\pi q_R^\pi \ket
+\tfrac{1}{4}F^\pi_{\mu\nu}F^\pi_{\mu\nu}
+\tfrac{1}{2}(\d_\mu A^\pi_\mu)^2
\\
&
\hspace*{1.5cm}
+ \tfrac{F_0^2}{8\Mk^2} (B_0/B)^2\, x_7
\Big] 
+ \ord{p^6}
\co
\\[1ex]
\bar{S}_4 &= 
F_0^2\int\dd^d x\,\Big\{
-4\Mk^2K_{8}\bra q_L^\pi q_R^\pi \ket
+(\tfrac{6}{5}K_{1} + \tfrac{1}{5}K_{2} + K_{5})\,w_{1}
+ (K_{2}+K_{6})\,w_{2}
\\
&
\hspace*{1.5cm}
+ K_{3}\,w_{3}
+ K_{4}\,w_{4}
+(\tfrac{6}{5}K_{7}+\tfrac{1}{5}K_{8}+\tfrac{4}{5}K_{9}-\tfrac{1}{5}K_{10}
)\,w_{5}
+(K_{8}+K_{10})\,w_{6}
\\
&
\hspace*{1.5cm}
+(K_{9}+K_{10})\,w_{7}
+K_{11}\,w_{8}
+K_{12}\,w_{9}
+K_{13}\,w_{10}
+K_{14}\,w_{11}
\\
&
\hspace*{1.5cm}
+F_0^2\big[(\tfrac{1}{25}K_{15}+\tfrac{6}{25}K_{16}+\tfrac{36}{25}K_{17})\,w_{12}
+(\tfrac{2}{5}K_{15}+\tfrac{6}{5}K_{16})\,w_{13}
+K_{15}\,w_{14}\big]
\Big\}
\\[0.5ex]
&+\bar{S}_4\big|_\mathrm{strong} 
+ \ord{p^6}
\fs
\end{split}
\end{equation}
%


\subsubsection{Loops: the determinant in the two--flavour limit}

The determinant of the differential operator $D$ covers all one--loop diagrams
of the generating functional. Its evaluation in terms of an expansion in
external fields in the two--flavour limit may be worked out as follows. To
begin with, we note that the contributions from the massless (pions 
and photon) and from the massive fields (kaons and eta) may be separated
\cite{Nyffeler,Schmid:PhD},  
\begin{equation}
\begin{split}
  \label{eq:det}
\ln\det D &= \ln\det D_{\ell}+\ln\det D_\eta+\ln\det D_{K}
+\ln\det(1-D_{\pi}^{-1}D_{\pi\eta}D_\eta^{-1} D_{\eta\pi})
\fs
\end{split}
\end{equation}
The first determinant $\ln\det D_\ell$ involves contributions from pions
and photons only, it is a purely non--local object and for the matching of the
\lecs{} needs not to be considered any further. The operators $D_\eta$, $D_K$
are related to heavy particles only: their determinant describe tadpoles with
insertions where only particles of identical masses run in the loop, either
etas or kaons, cf. Fig.~\ref{fig:det.mix} (left). Diagrams of this type are
efficiently calculated with heat--kernel methods. It
results in an expansion in terms of local quantities involving an increasing
number of derivatives, which corresponds to an expansion in powers of
momenta. The last determinant on the rhs. of Eq.~(\ref{eq:det}) is a more
complicated object. Since it involves the operator $D_\pi$ of the massless
modes, it is not a purely local object. However, its non--locality in the
low--energy expansion only shows up at order $p^6$ and can therefore be
neglected for our analysis. This is due to the symmetric operator
$D_{\pi\eta}$, which mediates $\pi-\eta$ mixing and is of order $p^2$. At
order $p^4$, the operators $D_\pi^{-1}$ and $D_\eta^{-1}$ may be replaced by
their free propagators. As a result, the diagram of Fig.~\ref{fig:det.mix}
(right) is the only contribution from the last term in Eq.~(\ref{eq:det}) at
order $p^4$. And the low--energy expansion of this diagram yields to local
terms only. The photon does not show up in this mixing term, because the eta
is not charged. We find
\begin{equation}
  \tfrac{1}{2}\ln(\det D/\det D^0)\big|_\mathrm{local} = 
\int\! \dd^dx\,
 \Big(\mathcal{L}^\eta_{1\,\mathrm{loop}}
+\mathcal{L}^K_{1\,\mathrm{loop}}
+\mathcal{L}^{\pi\eta}_{1\,\mathrm{loop}}\Big)+\ord{p^6}
\co
\end{equation}
\begin{equation}
\begin{split}
\label{eq:loops}
\mathcal{L}^\eta_{1\,\mathrm{loop}} 
&=
\tfrac{1}{24}F_1(M_\eta^2) 
\big[ \bra{\chi}_+^\pi\ket + M_K^{-2}\,x_7 \big]
  +\tfrac{1}{36}F_2(M_\eta^2)\,x_3
\co
\\[1.5ex]
\mathcal{L}^K_{1\,\mathrm{loop}} 
&=
-\tfrac{1}{4}F_1(M_K^2)
\big[ \bra{u}_\mu^\pi{u}_\mu^\pi-{\chi}^\pi_+\ket 
     -8\tfrac{C_0}{F_0^2}\bra q_R^\pi q_L^\pi \ket\big]
\\
&
\hspace*{-0.9cm}
+\tfrac{1}{48}F_2(M_K^2)\Big(x_1+2x_2+12x_4-2x_5-4x_6
-12x_7+x_9+24x_{10}-18x_{11}
\\
&
\hspace*{-0.9cm}
+\tfrac{C_0}{F_0^2}
\Big[\tfrac{48}{5}w_1+12w_2+\tfrac{12}{5}w_5+12w_6+36w_7
-6w_8
+\tfrac{C_0}{F_0^2}\big(
\tfrac{2496}{25}w_{12}+\tfrac{816}{5}w_{13}+48w_{14}\big)\Big]\Big)
\co
\\[1.5ex]
\mathcal{L}^{\pi\eta}_{1\,\mathrm{loop}} 
&=
 -\tfrac{1}{6}F_2^1(M^2_\eta)\big(x_7+x_8-x_{10}\big)
\co
\end{split}
\end{equation}
where $F_n^l(m^2)$ denote loop integrals, 
\begin{equation}
F_n^l(m^2)=
\int\!\frac{\dd^dq}{(2\pi)^d}\,\frac{1}{(m^2+q^2)^{n-l}(q^2)^l}
\co\qquad
n>l\geq0
\co\qquad
F_n(m^2) \equiv F_n^0(m^2)
\co
\end{equation}
which are well--known, cf. e.g. Ref.~\cite{GasserSainio}. 
The renormalisation is carried out in the $\overline{\mathrm{MS}}$--scheme,
where the 
\lecs{} $c_i\in\{l_i,k_i,L_i,K_i\}$ are splitted
into a divergent and a finite part as follows,
\begin{equation}
\begin{split}
c_i & = \alpha_i \lambda + c_i^\mathrm{r}(\mu,d)\co
\qquad
c_i^\mathrm{r}(\mu) \, \equiv \, c_i^\mathrm{r}(\mu,4)  \co\\[1ex]
\lambda &= \frac{\mu^{d-4}}{16\pi^2} 
\Big\{\frac{1}{d-4}-\tfrac{1}{2} [\ln 4\pi + \Gamma'(1) + 1] \Big\}.
\end{split}
\end{equation}
The coefficients $\alpha_i\in\{\gamma_i,\sigma_i,\Gamma_i,\Sigma_i\}$  are
given in Refs.~\cite{GL:Ann},\cite{Knecht.Urech},\cite{GL:NPB} and
\cite{Urech}\footnote{For $\Sigma_{15,16,17}$ consult
  Eq.~(12) in \cite{Steininger}.}.


\subsection{Results}

Collecting the results of the tree [Eq.~(\ref{eq:treeResults})] and one--loop
[Eq.~(\ref{eq:loops})] analysis allows us to determine
the relationship among the three-- and two--flavour
\lecs{} via Eq.~(\ref{eq:match_genF2}). The results are
%
\begin{align}
\label{eq:results}
   C&=C_0\left[1-4\mu_K\right]
 +4 M_K^2 F_0^2 \Kr{8} 
\co\nonumber
\\[0.5ex]
  \kr{1}&= \tfrac{6}{5}\Kr{1}
 +\tfrac{1}{5}\Kr{2}+\Kr{5}-\tfrac{2}{5}Z_0\nu_K
\co\nonumber
\\[0.5ex]
  \kr{2}&= \Kr{2}+\Kr{6}
-\tfrac{1}{2}Z_0\nu_K
\co\nonumber
\\[0.5ex]
  \kr{3}&= \Kr{3}
\co\nonumber
\\[0.5ex]
  \kr{4}&= \Kr{4}
\co\nonumber
\\[0.5ex]
  \kr{5}&= \tfrac{6}{5}K_7
+\tfrac{1}{5}\Kr{8}+\tfrac{4}{5}\Kr{9}-\tfrac{1}{5}\Kr{10}
-\tfrac{1}{10}Z_0\nu_K
\co\nonumber
\\[0.5ex]
  \kr{6}&= \Kr{8}+\Kr{10}
-\tfrac{1}{2}Z_0\nu_K
\co\nonumber
\\[0.5ex]
   k_7&= \Kr{9}+\Kr{10}
-\tfrac{3}{2}Z_0\nu_K
\co\nonumber
\\[0.5ex]
  \kr{8}&= \Kr{11}-2Z_0(2\Lr{4}
+\Lr{5})+\tfrac{1}{4}Z_0\nu_K
\co\nonumber
\\[0.5ex]
  \kr{9}&= \Kr{12}
\co\nonumber
\\[0.5ex]
   k_{10}&= K_{13}
\co\nonumber
\\[0.5ex]
   k_{11}&= K_{14}
\co\nonumber
\\[0.5ex]
  \kr{12}&= \tfrac{1}{25}\Kr{15}
+\tfrac{6}{25}\Kr{16}+\tfrac{36}{25}\Kr{17}-\tfrac{104}{25}Z_0^2\nu_K
\co\nonumber
\\[0.5ex]
  \kr{13}&= \tfrac{2}{5}\Kr{15}
+\tfrac{6}{5}\Kr{16}-\tfrac{34}{5}Z_0^2\nu_K
\co\nonumber
\\[0.5ex]
  \kr{14}&= \Kr{15}-2Z_0^2\nu_K
\co\nonumber
\\[0.5ex]
   l_{11}&=\tfrac{3}{2}\Lr{10}+3\Lr{11}
+\tfrac{3}{4}\nu_K 
\co
\end{align}
where we introduced the abbreviations, 
\begin{equation}
\begin{split}
  Z_0 &= C_0/F_0^4 \co \qquad
  \mu_K\,=\,\frac{M_K^2}{32\pi^2F_0^2}\ln\frac{M_K^2}{\mu^2}
\co \qquad
  \nu_K\,=\,\frac{1}{32\pi^2}\left(\ln\frac{M_K^2}{\mu^2}+1\right) 
\fs
\end{split}
\end{equation}
These relations are the main results of our article and deserve a few
comments:
\begin{itemize}
\item[--] 
  we only display the matching relations for
  the electromagnetic \lecs{} as well as the strong \lec{} $l_{11}$. The ones
  for the remaining strong \lecs{} may be established along the very same lines
  and may be found in \cite{GL:NPB,matchingI};
\item[--]
  at first glance it might come as a surprise that strong three--flavour
  \lecs{}  show up in the matching of the electromagnetic \lecs{},
  cf. $\kr{8}$. Along the derivation presented here it is the \eom{} which
  links strong and electromagnetic operators and needs to be used to project
  the operators $X_4$ and $X_5$ (in the two--flavour limit) into the
  two--flavour 
  basis; 
\item[--]
  briefly, we go back again to Eq.~(\ref{eq:loops}). We remark that
  only the kaon loop contributes non--trivially to the matching of the
  electromagnetic \lecs{}. Having in mind that along the method presented
  here, we had performed the matching for the strong \lecs{} $l_i$ at
  two--loops 
  already before \cite{matchingI,matchingII}, the relations in
  Eq.~(\ref{eq:results}) come at almost no additional cost. Exactly here lies the beauty
  of this approach: once the framework is set, all relations
  of the \lecs{} are obtained in one strike at a reasonable amount of effort;
\item[--] 
  one verifies that the dependence on the scale $\mu$ of the left and
  right hand side of Eq.~(\ref{eq:results}) is the same;
\item[--]
  specific linear combinations for electromagnetic matching relations were
  presented earlier in Ref.~\cite{Gasser:atom}. We completely agree with
  the relations given there. Some of the relations were also given in
  Ref.~\cite{Sazdjian-I}; see Ref.~\cite{Sazdjian-II} for comments why some
  of the relations in Ref.~\cite{Sazdjian-I} were given erroneously;
\item[--]
  the approach that was advocated in
  Refs.~\cite{Sazdjian-I,Gasser:atom,NehmePhD} relied on an analysis of
  physical 
  observables -- e.g. the charged pion mass or $\pi\pi$--scattering -- both in
  \cpt{3} as well as in \cpt{2}. The functional relationship among (linear
  combinations of) the \lecs{} emerges then from a comparison of the two
  representations in a large $m_s$ expansion. The agreement between the results
  found in the literature and those presented here based on the generating
  functional provides a thorough and welcome check on both calculations.
\end{itemize}


\section{Numerical analysis}
\label{sec:numerics}

The relations derived in Eq.~(\ref{eq:results}) are useful to obtain
constraints on the pertinent \lecs{}. In
the strong sector for instance, already in the early days of \chpt{} Gasser and
Leutwyler made use of such relations to convert information about the values
of the two--flavour \lecs{} to determine three--flavour \lecs{}; e.g.~an
estimate for $L_2$ was obtained in this way via $l_2$. 
The here presented formulae may by used in an analogous fashion: to the best
of our knowledge only estimates for the \lecs{} in \cpt{3} are known 
\cite{Anant.Mouss,Baur.Urech,Bijnens.Prades,MoussallamGauge,K_i}. The matching
relations may thus be inferred to obtain estimates on
the values of the two--flavour coupling constants. Two remarks shall be
pointed out which are relevant in the numerical determination of the \lecs{}:  
first, some of the \lecs{} depend on the gauge
\cite{BijnensGauge,MoussallamGauge}. The values inferred below are evaluated in
the Feynman gauge. Second, due to ultraviolet
divergences generated by photon loops, the splitting of the Hamiltonian of
$\text{\qcd{}} + \gamma$ into a strong and an electromagnetic piece is
ambiguous. This ambiguity must be reflected also in the effective theory in
the \lecs{} \cite{Scimemi}. Estimates of their sizes should therefore take
this into account. The authors of Ref.~\cite{Scimemi} have discussed the
problem in detail on the basis of fieldtheoretic models and come up with a
proposal how the ambiguity may be addressed systematically within these
models. Still, this delicate issue has not been investigated in the literature
for the \cpt{2,3} \lecs{} yet and is beyond the scope we are aiming at here. 

%
%
\begin{table}[t]
\begin{center}
\def\arraystretch{1.1}
\begin{tabular}{lrclr}
\hline\hline\\[-1ex]
$\Kr{1}$
&
$- 2.7$
&
\hspace*{1cm}
&
$K_7$
&
$0$
\\
$\Kr{2}$
&
$0.7$
&&
$\Kr{8}$
&
$0$
\\
$\Kr{3}$
&
$2.7$
&&
$\Kr{9}$
&
--
\\
$\Kr{4}$
&
$1.4$
&&
$\Kr{10}$
&
$4.0$
\\
$\Kr{5}$
&
$11.6$
&&
$\Kr{11}$
&
$1.3$
\\
$\Kr{6}$
&
$2.8$
&&
$\Kr{12}$
&
$-4.2$
\\
&&&
$K_{13}$
&
$4.7$
\\[1ex]
\hline\hline\\[-1ex]
\end{tabular}
\caption{
Values of electromagnetic {\lecs{}} in
  {\cpt{3}} in units of $10^{-3}$ at the scale $\mu=M_\rho=0.77\,\mathrm{GeV}$,
  in the Feynman gauge. The values for $\Kr{1},\ldots, \Kr{6}$ are
  invoked from \cite{Anant.Mouss}, and $K_7,\ldots,K_{13}$ from
  \cite{MoussallamGauge}, see
  also text for further details. 
}   
\label{tab:Kr}
\end{center}
\end{table}

%
%
\begin{table}[b]
\begin{center}
\def\arraystretch{1.1}
\begin{tabular}{lrclr}
\hline\hline\\[-1ex]
$\kr{1}$
&
$8.4$
&
\hspace*{1cm}
&
$\kr{6}$
&
$3.9$
\\
$\kr{2}$
&
$3.4$
&&
$k_7$
&
$3.7+\Kr{9}\cdot 10^3$
\\
$\kr{3}$
&
$2.7$
&&
$\kr{8}$
&
$-1.4$
\\
$\kr{4}$
&
$1.4$
&&
$\kr{9}$
&
$-4.2$
\\
$\kr{5}$
&
$-0.8+4/5\Kr{9}\cdot 10^3$
&&
$k_{10}$
&
$4.7$
\\[1ex]
\hline\hline\\[-1ex]
\end{tabular}
\caption{Values of electromagnetic low--energy constants in
  {\cpt{2}} in 
  units of $10^{-3}$ at the scale $\mu=M_\rho=0.77\,\mathrm{GeV}$, in the
  Feynman gauge. From general dimensional arguments, one might attribute an
  uncertainty of $1/(16\pi^2)\approx 6.3\cdot 10^{-3}$ to each \lec{}.}
\label{tab:kr}
\end{center}
\end{table}

\begin{center}
\begin{figure}[t]
\begin{minipage}[]{1.0\linewidth}
\epsfig{file=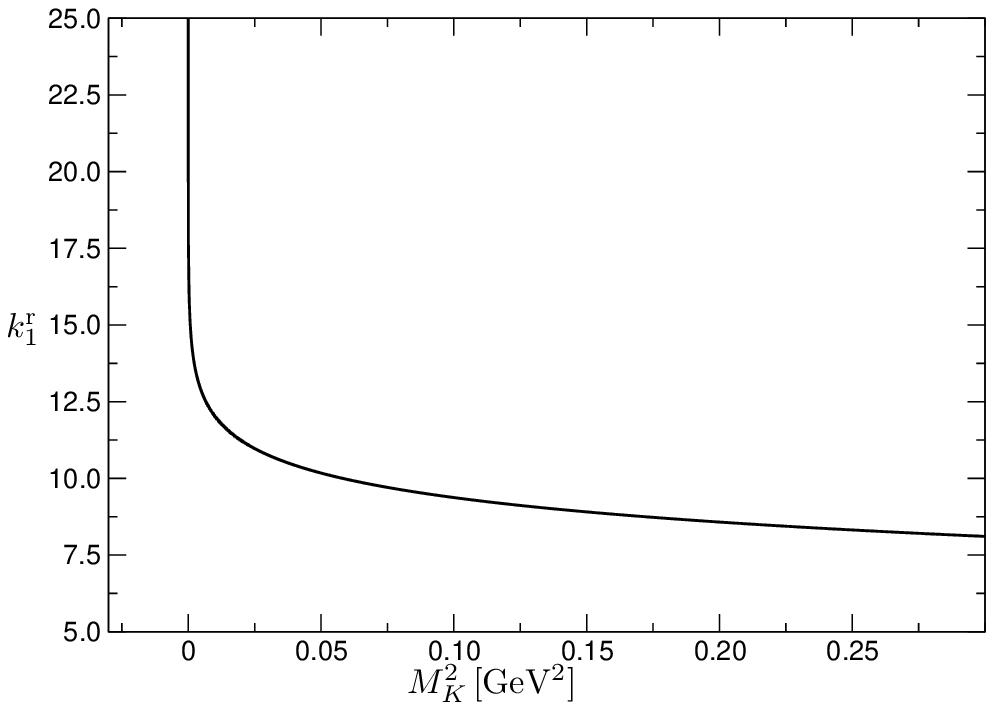,width=0.48\linewidth}
\hfill
\epsfig{file=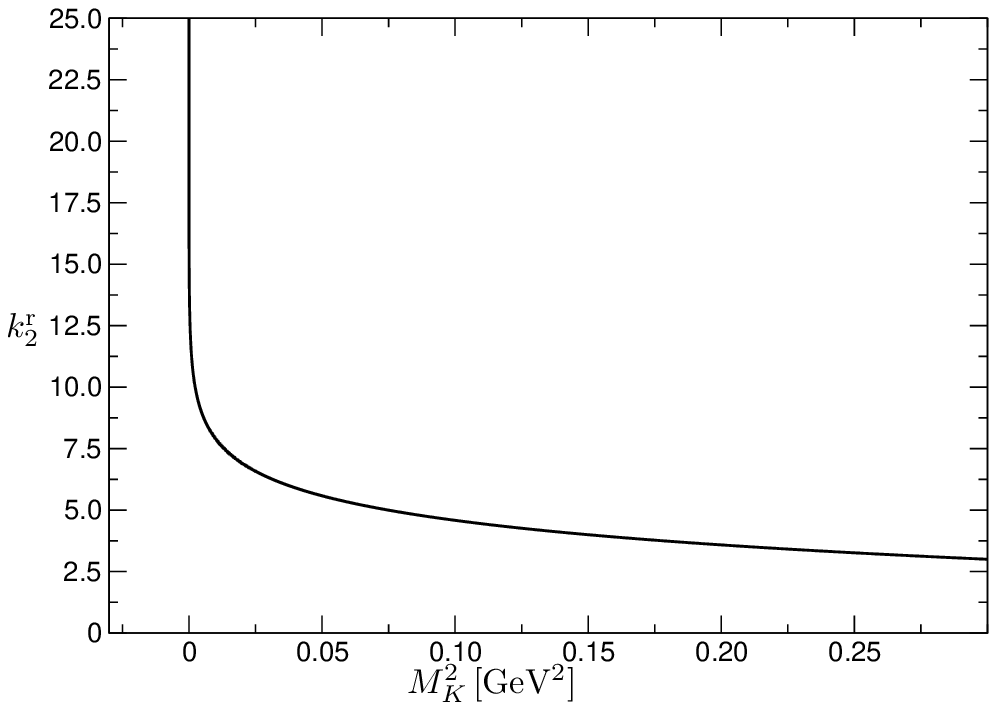,width=0.48\linewidth}\\
\end{minipage}
\caption{Strange quark mass dependence of the electromagnetic two--flavour
  {\lecs{}} $\kr{1}$ (left) and $\kr{2}$ (right) in units of $10^{-3}$  at the
  scale $\mu=M_\rho=0.77\,\mathrm{GeV}$. The physical value of $m_s$ 
  corresponds to $M_K^2 = B_0m_s \approx (485\,\mathrm{MeV})^2$. 
}
\label{fig:kr1kr2.eps}       
\end{figure}
\end{center}

We proceed with the numerical analysis of Eq.~(\ref{eq:results}). At the
accuracy we are working, we may identify $F_0$ with the pion decay constant
$F_\pi=92.4\,\mathrm{MeV}$ and  $M_K^2 = B_0 m_s$ with
\begin{equation}
  B_0 m_s = M_{K^+}^2 - M_{\pi^+}^2/2 \simeq (485 \,\mathrm{MeV})^2
\fs
\end{equation}
We further set $Z_0 = 0.91$ (obtained from Ref.~\cite[Eq.~(42)]{MoussallamGauge}
with $M_V = 0.77\,\mathrm{GeV}$, $z\equiv M_A^2/M_V^2 = 2$). For the values of
the three--flavour \nlo{} electromagnetic \lecs{} we will stick
 for $\Kr{1},\ldots,\Kr{6}$ to \cite{Anant.Mouss}, and  for
$K_7,\ldots,\Kr{12}$ to \cite{MoussallamGauge}, summarised here in 
Tab.~\ref{tab:Kr}. The \lec{} $\Kr{9}$
remained undetermined in \cite{MoussallamGauge}, as it yet suffers from a
reliable estimate (consult
\cite{MoussallamGauge} for the details). As a result we will not give a
numerical estimate for $\kr{5}$ and $k_7$. For the \lecs{} $\Kr{11},\Kr{12}$
(from Eq.(59) and (61) in \cite{MoussallamGauge}) we furthermore set $\mu_0 =
1\,\mathrm{GeV}$ for the \qcd{} scale, the parameters $M_V$ and $z$ are as
introduced above. The coupling constants $\Kr{14},\ldots,\Kr{17}$ are
associated to contact operators and/or operators at order $e^4$ and are not
considered in this section. Furthermore, $\Lr{4} = 0$ and $\Lr{5} = 1.5\cdot
10^{-3}$, taken from the $\ord{p^4}$ fit in \cite{BijnensFit}. All \lecs{} are
evaluated at the scale $\mu=0.77\,\mathrm{GeV}$, and in the Feynman gauge
$\xi=1$. The results for the so obtained electromagnetic two--flavour \lecs{} 
are finally summarised in Tab.~\ref{tab:kr}. As an illustration we also show in
Fig.~\ref{fig:kr1kr2.eps} the strange quark mass dependence of $\kr{1}$ and
$\kr{2}$. We observe that the two--flavour \lecs{} only show 
a very moderate strange quark mass dependence in the neighbourhood of the
physical point. This pattern is due to its solely logarithmic strange quark
mass dependence at this order of the matching. Note 
that the matching relations only apply over a certain range for
the strange quark mass: the formulae break down for $m_s \rightarrow 0$, as
the expansion performed here requires that all external momenta are much
smaller than $m_s$. Remarkably, the pertinent chiral logarithm becomes dominant
numerically only for very small $m_s$. 
On the other hand as one increases the strange
quark mass, higher order contributions in the matching expansion become more
dominant and start to spoil the behaviour of the chiral logarithm. This was
discussed in more detail in \cite{matchingI} for the strong \lecs{} $l_i$. 

%
%

We consider it difficult to assign reliable errors to the estimates of the
\lecs{} in Tab.~\ref{tab:kr}. The determinations in
\cite{MoussallamGauge,Anant.Mouss,Baur.Urech} -- from where we invoked the
$\Kr{i}$ -- are model dependent, for which reliable estimates of uncertainties
are always a delicate affair. Moreover, the scale dependence in various
\lecs{} can be strongly correlated. We shall therefore refrain from assigning
individual errors to the estimates in Tab.~\ref{tab:kr}. To be conservative,
one might attribute an uncertainty of $1/(16\pi^2)\approx 6.3\cdot 10^{-3}$ to
each \lec{} $\kr{i}$, stemming from general dimensional arguments. The size of
this uncertainty compared to the values in Tab.~\ref{tab:kr} indicates that
the entries of the table 
are yet only a rough order of magnitude estimate.  

In the near future, a more precise determination of (some combinations of)
electromagnetic \lecs{} may also be expected from lattice
\qcd{}. In this respect we mention two recent studies that address the
electromagnetic splitting of pseudoscalar meson masses \cite{RBC,JLQCD}.


\section{Conclusions}
\label{sec:conclusions}

In summary we have worked out the strange quark mass dependence of the
two--flavour electromagnetic \lecs{} $C,k_i$ at next--to--leading order. The
calculation relied on a non--trivial matching between the three--flavour 
and the two--flavour generating functional of \chpt{} 
including virtual photons and amounts to 16 relations among the \lecs{} of
\cpt{2} and \cpt{3}.\\  

\noindent
These relations are useful to obtain constraints and further information on
the pertinent \lecs{}. As an application we have used these relations to obtain
numerical estimates for the values of the two--flavour electromagnetic
\lecs{}.


\subsection*{Acknowledgements}

We thank J\"urg Gasser for suggesting us to undertake this work, and are
grateful for many enlightening discussions and comments on the
manuscript. We further thank Marc Knecht, Bachir Moussallam and Hagop Sazdjian
for useful discussions and/or comments on the manuscript. C.H. and M.A.I. wish
to express their gratitude to the members of the Institute of Theoretical
Physics, University of Bern, for their kind hospitality during the final stages of
this work. This work was supported  by the Swiss National Science Foundation,
by the Ministerio de Educaci\'on y Ciencia under the project FPA2004-00996, by
Generalitat Valenciana GVACOMP2007-156, and by EU MRTN-CT-2006-035482
(FLAVIA{\it net}).



\begin{thebibliography}{99}
\setlength{\baselineskip}{14pt}

\bibitem{Weinberg}
  S.~Weinberg,
  Physica A {\bf 96} (1979) 327.

\bibitem{GL:Ann}
  J.~Gasser and H.~Leutwyler,
  Annals Phys.\  {\bf 158} (1984) 142.

\bibitem{GL:NPB}
  J.~Gasser and H.~Leutwyler,
  Nucl.\ Phys.\  B {\bf 250} (1985) 465.

\bibitem{Ecker:lecs}
  G.~Ecker,
  arXiv:hep-ph/0702263.

\bibitem{Bijnens:lecs}
  J.~Bijnens,
  arXiv:0708.1377 [hep-lat].

\bibitem{Urech}
  R.~Urech,
  Nucl.\ Phys.\  B {\bf 433} (1995) 234
  [hep-ph/9405341].

\bibitem{Neufeld}
  H.~Neufeld and H.~Rupertsberger,
  Z.\ Phys.\  C {\bf 71} (1996) 131
  [hep-ph/9506448];
  H.~Neufeld and H.~Rupertsberger,
  Z.\ Phys.\  C {\bf 68}, 91 (1995).

\bibitem{Knecht.Urech}
  M.~Knecht and R.~Urech,
  Nucl.\ Phys.\  B {\bf 519} (1998) 329
  [hep-ph/9709348].

\bibitem{Steininger}
  U.~G.~Meissner, G.~Muller and S.~Steininger,
  Phys.\ Lett.\  B {\bf 406} (1997) 154
  [Erratum-ibid.\  B {\bf 407} (1997) 454]
  [hep-ph/9704377].

\bibitem{Anant.Mouss}
  B.~Ananthanarayan and B.~Moussallam,
  JHEP {\bf 0406} (2004) 047
  [hep-ph/0405206].

\bibitem{Baur.Urech}
  R.~Baur and R.~Urech,
  Nucl.\ Phys.\  B {\bf 499}, 319 (1997)
  [hep-ph/9612328].

\bibitem{Bijnens.Prades}
  J.~Bijnens and J.~Prades,
  Nucl.\ Phys.\  B {\bf 490}, 239 (1997)
  [hep-ph/9610360];

\bibitem{MoussallamGauge}
  B.~Moussallam,
  Nucl.\ Phys.\  B {\bf 504}, 381 (1997)
  [hep-ph/9701400];

\bibitem{K_i}
  V.~E.~Lyubovitskij, T.~Gutsche, A.~Faessler and R.~Vinh Mau,
  Phys.\ Lett.\  B {\bf 520}, 204 (2001)
  [hep-ph/0108134];
  V.~E.~Lyubovitskij, T.~Gutsche, A.~Faessler and R.~Vinh Mau,
  Phys.\ Rev.\  C {\bf 65}, 025202 (2002)
  [hep-ph/0109213].

\bibitem{Gasser:atom}
  J.~Gasser, V.~E.~Lyubovitskij, A.~Rusetsky and A.~Gall,
  Phys.\ Rev.\  D {\bf 64} (2001) 016008
  [hep-ph/0103157].

\bibitem{Sazdjian-I}
  H.~Jallouli and H.~Sazdjian,
  Phys.\ Rev.\  D {\bf 58}, 014011 (1998)
  [Erratum-ibid.\  D {\bf 58}, 099901 (1998)]
  [hep-ph/9706450].

\bibitem{Nyffeler}
  A.~Nyffeler and A.~Schenk,
  Annals Phys.\  {\bf 241} (1995) 301
  [hep-ph/9409436].

\bibitem{Schmid:PhD}
M.~Schmid, {\it Strangeless $\chi PT$ at large $m_s$}, PhD thesis, 
University of  Bern, January 2007.

\bibitem{matchingI}
  J.~Gasser, C.~Haefeli, M.~A.~Ivanov and M.~Schmid,
  Phys.\ Lett.\  B {\bf 652} (2007) 21
  [arXiv:0706.0955 [hep-ph]].

\bibitem{NehmePhD}
  A.~Nehme, {\it ``La Brisure d'Isospin dans les Interactions M\'eson--M\'eson \`a
    Basse Energie''}, PhD thesis, CPT Marseille, July 2002.

\bibitem{Mouss:Sigma}
  B.~Moussallam,
  JHEP {\bf 0008}, 005 (2000)
  [hep-ph/0005245].

\bibitem{Schweizer:B}
  R.~Kaiser and J.~Schweizer,
  JHEP {\bf 0606}, 009 (2006)
  [hep-ph/0603153].

\bibitem{meissner_frink}
  M.~Frink and U.~G.~Meissner,
  JHEP {\bf 0407}, 028 (2004)
  [hep-lat/0404018].

\bibitem{WZW}
  J.~Wess and B.~Zumino,
  Phys.\ Lett.\  B {\bf 37} (1971) 95;
  E.~Witten,
  Nucl.\ Phys.\  B {\bf 223} (1983) 422.

\bibitem{Kaiser}
  R.~Kaiser,
  Phys.~Rev.~{\bf D63}, 76010 (2001)
  [hep-ph/0011377].

\bibitem{MoussAnomaly}
  B.~Ananthanarayan and B.~Moussallam,
  JHEP {\bf 0205} (2002) 052
  [hep-ph/0205232].

\bibitem{Schweizer}
  J.~Schweizer,
  JHEP {\bf 0302}, 007 (2003)
  [hep-ph/0212188].

\bibitem{matchingII}
  J.~Gasser, C.~Haefeli, M.~A.~Ivanov and M.~Schmid,
  in preparation.

\bibitem{GasserSainio}
    J.~Gasser and M.~E.~Sainio,
  Eur.\ Phys.\ J.\  C {\bf 6} (1999) 297
  [hep-ph/9803251].

\bibitem{Sazdjian-II}
  H.~Sazdjian,
  Phys.\ Lett.\  B {\bf 490}, 203 (2000)
  [hep-ph/0004226].

\bibitem{BijnensGauge}
  J.~Bijnens,
  Phys.\ Lett.\  B {\bf 306}, 343 (1993)
  [hep-ph/9302217];

\bibitem{Scimemi}
  J.~Gasser, A.~Rusetsky and I.~Scimemi,
  Eur.\ Phys.\ J.\  C {\bf 32} (2003) 97
  [hep-ph/0305260].

\bibitem{BijnensFit}
  G.~Amoros, J.~Bijnens and P.~Talavera,
  Nucl.\ Phys.\  B {\bf 602}, 87 (2001)
  [hep-ph/0101127].

\bibitem{RBC}
  T.~Blum, T.~Doi, M.~Hayakawa, T.~Izubuchi and N.~Yamada,
  arXiv:0708.0484 [hep-lat].

\bibitem{JLQCD}
  E.~Shintani, H.~Fukaya, S.~Hashimoto, H.~Matsufuru, J.~Noaki, T.~Onogi and
  N.~Yamada 
  [JLQCD Collaboration],
  arXiv:0710.0691 [hep-lat].


\end{thebibliography}
\end{document}